\documentclass[12pt]{article}

\def\be{\begin{equation}}
\def\ee{\end{equation}}
\def\ba{\begin{array}{c}}
\def\ea{\end{array}}
\def\bea{\begin{eqnarray}}
\def\eea{\end{eqnarray}}

\def\ben{\begin{displaymath}}
\def\een{\end{displaymath}}

\newcommand{\bas}{\begin{eqnarray*}}
\newcommand{\eas}{\end{eqnarray*}}

\newcommand{\sech}{\mathop{\rm sech}\nolimits}
\newcommand{\csch}{\mathop{\rm csch}\nolimits}
\newcommand{\ca}{{\cal A}}
\newcommand{\bca}{\bar{\cal A}}

\title{\boldmath${\cal PT}$-supersymmetric partner of a short-range
 square well
       }
\author{C Quesne$^a$, B Bagchi$^b$, H B\'\i la$^c$, V Jakubsk\'y$^c$, S Mallik$^b$, 
M Znojil$^c$\\
{\small $^c$ Physique Nucl\'eaire Th\'eorique et Physique Math\'ematique,} \\
{\small Universit\'e Libre de Bruxelles, Campus de la Plaine CP229,} \\
{\small Boulevard~du
Triomphe, B-1050 Brussels, Belgium} \\
{\small $^b$ Department of Applied Mathematics, University of Calcutta,} \\
{\small 92 Acharya Prafulla Chandra Road, Kolkata 700 009, West Bengal, India}\\
{\small $^c$ \'Ustav jadern\'e fyziky AV \v CR, 250 68 \v Re\v z, Czech Republic}}
\date{ }
\begin{document}
\baselineskip=22pt plus 1pt minus 1pt
\maketitle
\begin{abstract} In a box of size $L$, a spatially antisymmetric square-well potential of a
purely imaginary strength ${\rm i}g$ and size
$l < L$ is interpreted as an initial element of the SUSY hierarchy of solvable
Hamiltonians, the energies of which are all real for $g < g_c(l)$. The first
partner potential is constructed in closed form and discussed.
\end{abstract}

\section{Introduction}     

The technically slightly complicated but quantum-mechanically straightforward solution of
the one-dimensional, ${\cal PT}$-symmetric Schr\"{o}dinger equation
 \be
  \left(- \frac{d^2}{d x^2}+V(x)\right)\psi_n(x)
  =E_n\psi_n(x), \qquad n = 0, 1, 2, \ldots,
  \label{eq:problem}
 \ee 
with the Dirichlet boundary conditions $\psi(\pm L)=0$ and with a purely imaginary
$V(x)$ may be found elsewhere
\cite{Hynek,Langer,sqw}. Here, such a model with real spectrum and
\be
  V(x)=V^{(+)}(x)=\left\{\begin{array}{ll}
       0 & {\rm for\ } L > |x| > l, \\[0.1cm]
       {\rm i}g\, {\rm sign}\, x, \quad g > 0, & {\rm for\ } |x| \le l,
 \end{array}\right.  \label{eq:V-i}
\ee
will be considered factorized and complemented by another similar model,
\be
   - \frac{d^2}{dx^2} + V^{(+)}(x) - D_0= \bca \ca \equiv H^{(+)}, \quad
   \ca \bca = - \frac{d^2}{dx^2} + V^{(-)}(x) - D_0 \equiv H^{(-)},
  \label{eq:Hpm}
\ee
where the well-known operators and identities
\be
  \ca = \frac{d}{dx} + W(x), \qquad \bca = - \frac{d}{dx} + W(x),
  \qquad  V^{(\pm)}- D_0 = W^2 \mp W'  \label{eq:definition}
 \ee
are employed. We denote the wave functions of $H^{(+)}$ (resp.\ $H^{(-)}$) by symbols
$\psi^{(+)}_n(x)$ (resp.\ $\psi^{(-)}_n(x)$), $n=0$, 1, 2,~\ldots, and we assume
the so-called unbroken-supersymmetry condition $ \ca\, \psi^{(+)}_0(x) = 0 $ of
Witten's supersymmetric quantum mechanics (SUSYQM) \cite{SUSYQM} (hence $D_0 =
E^{(+)}_0$ hereabove). As long as the application of such a formalism to non-Hermitian
operators is always subject to caution, we believe that both the construction and some
unusual properties of the partner potential $V^{(-)}(x)$ deserve an explicit description.
%
%
\section{\boldmath The ${\cal PT}$-symmetric SUSY partner potential $V^{(-)}(x)$
\label{sec:SUSYQM}} 

The purpose of the present section is to construct and study the
SUSY partner  $H^{(-)}$ of the square-well Hamiltonian $H^{(+)}$
in the physically-relevant unbroken $\cal PT$-symmetry regime,
corresponding to $g < g_c(l)$ of ref. \cite{Hynek}.
%
%
\subsection{Determination of the parameters \label{sec:W}}

Let us denote the four regions $-L < x < -l$, $-l < x < 0$, $0 < x
< l$, $l < x < L$ by $L2$, $L1$, $R1$, $R2$, respectively, and write for
$V^{(+)}$, defined in (\ref{eq:V-i}),  $V^{(+)}_{L2}(x) = 0$, $V^{(+)}_{L1}(x) = - {\rm
i} g$, $V^{(+)}_{R1}(x) = {\rm i} g$, $V^{(+)}_{R2}(x) = 0$. Setting
$D_0 = E^{(+)}_0$, where
\be
  E^{(+)}_n = k_n^2 = t_n^2 - s_n^2, \qquad \kappa_n = s_n + {\rm i} t_n, \qquad
  g = 2 s_n t_n,  \label{eq:energy}
\ee
for $n=0$, 1, 2,~\ldots, we obtain for the superpotential and the partner potential the
respective formulae
\be
  W(x) = \left\{\begin{array}{l}
      W_{L2}(x) = k_0 \tan[k_0(x + x_{L2})] \\[0.1cm]
      W_{L1}(x) = - \kappa_0^* \tanh[\kappa_0^*(x + x_{L1})] \\[0.1cm]
      W_{R1}(x) = - \kappa_0 \tanh[\kappa_0(x - x_{R1})] \\[0.1cm]
      W_{R2}(x) = k_0 \tan[k_0(x - x_{R2})]
  \end{array} \right.
\ee
and
\be
  V^{(-)}(x) = \left\{\begin{array}{l}
      V^{(-)}_{L2}(x) = 2 k_0^2 \sec^2[k_0(x + x_{L2})] \\[0.1cm]
      V^{(-)}_{L1}(x) = - 2 \kappa_0^{*2} \sech^2[\kappa_0^*(x
       + x_{L1})] - {\rm i} g
           \\[0.1cm]
      V^{(-)}_{R1}(x) = - 2 \kappa_0^2 \sech^2[\kappa_0(x
      - x_{R1})] + {\rm i} g  \\[0.1cm]
      V^{(-)}_{R2}(x) = 2 k_0^2 \sec^2[k_0(x - x_{R2})]
  \end{array} \right..
  \label{eq:partner-0}
\ee
Here $x_{L2}$, $x_{L1}$, $x_{R1}$ and $x_{R2}$ denote four
integration constants. We choose
\be
  x_{L2} = L + \frac{\pi}{2k_0}, \qquad x_{R2} = L - \frac{\pi}{2k_0}
   \label{eq:integration-2}
\ee
to ensure that $V^{(-)}_{L2}$ and $V^{(-)}_{R2}$ blow up at the
end points $x=-L$ and $x=L$. This is in tune with~\cite{CQ}. We
thus get
\be
  V^{(-)}_{L2}(x) = 2 k_0^2 \csc^2[k_0(x + L)],
   \qquad V^{(-)}_{R2}(x) = 2 k_0^2
  \csc^2[k_0(x - L)]. \label{eq:partner-bis}
\ee
Observe that for the superpotential, $W_{L2}(x)$ and $W_{R2}(x)$
also blow up at these points:
\be
  W_{L2}(x) = - k_0 \cot[k_0(x + L)], \qquad W_{R2}(x)
   = - k_0 \cot[k_0(x - L)].
 \ee
\par
%
%
The ground-state wavefunction of $H^{(+)}$ is given by~\cite{Hynek}
\bea
  \psi^{(+)}_{0R2}(x) & = & \psi^{(+)*}_{0L2}(-x)
  = A^{(+)}_0 \sin[k_0(L - x)], \\
  \psi^{(+)}_{0R1}(x) & = & \psi^{(+)*}_{0L1}(-x)
   = B^{(+)}_0 \cosh(\kappa_0 x) + {\rm i}
       \frac{C^{(+)}_0}{\kappa_0 l} \sinh(\kappa_0 x),
\eea
where $A^{(+)}_0$, $B^{(+)}_0$, $C^{(+)}_0$ are three constants, $B^{(+)}_0$,
$C^{(+)}_0$ are real and
\bea
  A^{(+)}_0 & = & B^{(+)}_0\, \frac{\kappa_0 \csc[k_0(L-l)] \csch(\kappa_0 l)}
         {k_0 \cot[k_0(L-l)] + \kappa_0 \coth(\kappa_0 l)}, \label{eq:A} \\
  C^{(+)}_0 & = & {\rm i} \kappa_0 l B^{(+)}_0\, \frac{k_0 \cot[k_0(L-l)] 
         \coth(\kappa_0 l) + \kappa_0}{k_0 \cot[k_0(L-l)] + \kappa_0 \coth(\kappa_0 l)},
         \label{eq:C}
\eea
as a result of the matching conditions on $\psi^{(+)}_0(x)$ and its derivative at $x = 0$
and $x = \pm l$. It turns out that the unbroken-SUSY condition is automatically
satisfied in the regions $R2$ and $L2$ due to the choice made for the integration
constants $x_{R2}$, $x_{L2}$ in (\ref{eq:integration-2}). In the region $R1$, we find a
condition fixing the value of $x_{R1}$,
\be
  \tanh(\kappa_0 x_{R1}) = - \frac{{\rm i} C^{(+)}_0}{\kappa_0 l
  B^{(+)}_0}
   = \frac{k_0
  \cot[k_0(L-l)] \coth(\kappa_0 l) + \kappa_0}{k_0 \cot[k_0(L-l)]
   + \kappa_0
  \coth(\kappa_0 l)} \label{eq:x_R1}.
\ee
A similar relation applies in $L1$, thus leading to the result
\be
  x_{L1} = x_{R1}^*. \label{eq:x_L1}
 \ee
Note that in contrast with the real integration constants
$x_{R2}$, $x_{L2}$, the constants $x_{R1}$ and $x_{L1}$ are
complex. Separating both sides of equation (\ref{eq:x_R1}) into a
real and an imaginary part, we obtain the two equations
\bea
  \frac{\sinh X \cosh X}{\cosh^2 X \cos^2 Y + \sinh^2 X \sin^2 Y} &
  = & \frac{N^r}{D},
       \label{eq:x_R1-1} \\
  \frac{\sin Y \cos Y}{\cosh^2 X \cos^2 Y + \sinh^2 X \sin^2 Y} &
  = & \frac{N^i}{D},
       \label{eq:x_R1-2}
\eea
where we have used the decompositions $\kappa_0 = s_0 + {\rm
i}t_0$, $x_{R1} = x_{R1}^r + {\rm i} x_{R1}^i$, $\kappa_0 x_{R1} =
X + {\rm i} Y$, implying that
\be
  X = s_0 x_{R1}^r - t_0 x_{R1}^i, \qquad Y = t_0 x_{R1}^r
  + s_0 x_{R1}^i,
\ee
and we have defined
\be
  N^r  = \{- s_0^2 \cos[2k_0(L-l)] + t_0^2\} \sinh(2s_0 l)
  + k_0 s_0 \sin[2k_0(L-l)]
       \cosh(2s_0 l),
\ee
\be
  N^i  = \{s_0^2 - t_0^2  \cos[2k_0(L-l)]\} \sin(2t_0 l)
   - k_0 t_0 \sin[2k_0(L-l)]
       \cos(2t_0 l),
\ee
\bea
  D & = & \{- s_0^2 \cos[2k_0(L-l)] + t_0^2\} \cosh(2s_0 l)
  + \{s_0^2 - t_0^2
        \cos[2k_0(L-l)]\} \cos(2t_0 l) \nonumber \\
  && \mbox{} + k_0 \sin[2k_0(L-l)] [s_0 \sinh(2s_0 l)
   + t_0 \sin(2t_0 l)].
\eea
Equations (\ref{eq:x_R1-1}) and (\ref{eq:x_R1-2}), when solved
numerically, furnish the values of both the parameters $x_{R1}^r$
and $x_{R1}^i$. One may also observe that the resulting
superpotential $W(x) = - W^*(-x)$ and partner potential
$V^{(-)}(x) = V^{(-)*}(-x)$ are $\cal PT$-antisymmetric and $\cal
PT$-symmetric, respectively.\par
%
%
\subsection{Eigenfunctions in the partner potential
 \label{sec:eigenfunctions}}

On exploiting the SUSY intertwining relations, the eigenfunctions
$\psi^{(-)}_n(x)$, $n=0$, 1, 2,~\ldots, of $H^{(-)}$ can be
obtained by acting with $\ca$ on $\psi^{(+)}_{n+1}(x)$, subject to
the preservation of the boundary and continuity conditions
\bea
  \psi^{(-)}_{nL2}(-L) & = & 0, \qquad \psi^{(-)}_{nR2}(L) = 0,
   \label{eq:boundary} \\
  \psi^{(-)}_{nL2}(-l) & = & \psi^{(-)}_{nL1}(-l), \qquad
   \partial_x \psi^{(-)}_{nL2}(-l) =
        \partial_x \psi^{(-)}_{nL1}(-l), \label{eq:continuity-1}
         \\
  \psi^{(-)}_{nL1}(0) & = & \psi^{(-)}_{nR1}(0), \qquad
   \partial_x \psi^{(-)}_{nL1}(0) =
        \partial_x \psi^{(-)}_{nR1}(0), \label{eq:continuity-2}
         \\
  \psi^{(-)}_{nR1}(l) & = & \psi^{(-)}_{nR2}(l), \qquad
   \partial_x \psi^{(-)}_{nR1}(l) =
        \partial_x \psi^{(-)}_{nR2}(l). \label{eq:continuity-3}
\eea
\par
%
%
Application of $\ca$ leads to the forms
\bea
  \psi^{(-)}_{nL2}(x) & = & C^{(-)}_{nL2}\,
  A^{(+)*}_{n+1} \sin[k_{n+1}(L+x)]\nonumber \\
  && \mbox{} \times \{k_{n+1} \cot[k_{n+1}(L+x)]
   - k_0 \cot[k_0(L+x)]\},
        \label{eq:partner-psi-1} \\
  \psi^{(-)}_{nL1}(x) & = & C^{(-)}_{nL1}\, B^{(+)}_{n+1}
   \sinh(\kappa_{n+1}^* x)
        \{\kappa_{n+1}^* - \kappa_0^* \tanh[\kappa_0^*(x
        + x_{R1}^*)]
        \coth(\kappa_{n+1}^* x)\} \nonumber \\
  && \mbox{} + C^{(-)}_{nL1}\, \frac{{\rm i}
  C^{(+)}_{n+1}}{\kappa_{n+1}^* l}
        \sinh(\kappa_{n+1}^* x) \nonumber \\
  && \mbox{} \times \{\kappa_{n+1}^* \coth(\kappa_{n+1}^* x)
  - \kappa_0^*
        \tanh[\kappa_0^*(x + x_{R1}^*)]\}, \\
  \psi^{(-)}_{nR1}(x) & = & C^{(-)}_{nR1}\, B^{(+)}_{n+1}
   \sinh(\kappa_{n+1} x)
        \{\kappa_{n+1} - \kappa_0 \tanh[\kappa_0(x - x_{R1})]
        \coth(\kappa_{n+1} x)\} \nonumber \\
  && \mbox{} + C^{(-)}_{nR1}\,
  \frac{{\rm i} C^{(+)}_{n+1}}{\kappa_{n+1} l}
        \sinh(\kappa_{n+1} x) \nonumber \\
  && \mbox{} \times \{\kappa_{n+1} \coth(\kappa_{n+1} x)
   - \kappa_0
        \tanh[\kappa_0(x - x_{R1})]\}, \\
  \psi^{(-)}_{nR2}(x) & = & C^{(-)}_{nR2}\, A^{(+)}_{n+1}
   \sin[k_{n+1}(L-x)]\nonumber \\
  && \mbox{} \times \{- k_{n+1} \cot[k_{n+1}(L-x)]
  + k_0 \cot[k_0(L-x)]\},
        \label{eq:partner-psi-4}
\eea
where $C^{(-)}_{nL2}$, $C^{(-)}_{nL1}$, $C^{(-)}_{nR1}$,
$C^{(-)}_{nR2}$ denote some complex constants and equation
(\ref{eq:x_L1}) has been used. Boundary conditions
(\ref{eq:boundary}) are satisfied. It remains to impose the
continuity conditions (\ref{eq:continuity-1}) --
(\ref{eq:continuity-3}).\par
%
%
The matching of the regions $L1$ and $R1$ at $x=0$ leads to two conditions, which are
compatible because the two constraints
\bea
  \kappa_0 \tanh(\kappa_0 x_{R1}) & = &
  - \kappa_0^* \tanh(\kappa_0^* x_{R1}^*), \label{eq:C1}\\
  \kappa_{n+1}^{*2} - \kappa_{n+1}^2 & = & \kappa_0^{*2}
   - \kappa_0^2 = - 2{\rm i}g,  \label{eq:C2}
 \eea
are satisfied owing to (\ref{eq:x_R1}) and (\ref{eq:energy}), respectively. It therefore
remains a single condition
\be
  C^{(-)}_{nR1} = C^{(-)}_{nL1}.
\ee
For the matching between $R1$ and $R2$ at $x=l$, a similar situation happens due this
time to the two constraints
\bea
  \kappa_0 \tanh[\kappa_0(l - x_{R1})] & = & - k_0 \cot[k_0(L-l)],
      \label{eq:relation-1} \\
  \kappa_{n+1}^2 - \kappa_0^2 & = & k_0^2 - k_{n+1}^2.
\eea
The resulting condition reads
\be
  C^{(-)}_{nR1} = C^{(-)}_{nR2}. \label{eq:C-R12}
\ee
Since a result similar to (\ref{eq:C-R12}) applies at the
interface between regions $L2$ and $L1$, we conclude that the
partner potential eigenfunctions are given by equations
(\ref{eq:partner-psi-1}) -- (\ref{eq:partner-psi-4}) with
\be
  C^{(-)}_{nL2} = C^{(-)}_{nL1} = C^{(-)}_{nR1} = C^{(-)}_{nR2}
   \equiv C^{(-)}_n. \ee
Such eigenfunctions are $\cal PT$-symmetric provided we choose
$C^{(-)}_n$ imaginary:
\be
  C^{(-)*}_n = - C^{(-)}_n.
\ee
\par
%
%
\section{Discontinuities in the partner potential $V^{(-)}(x)$
\label{sec:discontinuities}}

In subsection \ref{sec:W}, we have constructed the SUSY partner
$V^{(-)}(x)$ of a piece-wise potential with three discontinuities
at $x^{(i)} = -l$, 0 and $l$, where $i=1$, 2, 3. We may now ask the following question:
does the former have the same discontinuities as the latter or could the discontinuity
number decrease? We plan to prove here that the second alternative can be ruled out.\par
%
%
{}For such a purpose, let us determine the jump (if any) of the partner potential at
$x^{(i)}$, $\Delta V^{(-)}(x^{(i)}) \equiv \lim_{x\to x^{(i)}_+} V^{(-)}(x) -
\lim_{x\to x^{(i)}_-} V^{(-)}(x)$. A simple calculation leads to $\Delta V^{(-)}(0) =
- 2 \kappa_0^2 \sech^2(\kappa_0 x_{R1}) + {\rm i}g - [- 2 \kappa^{*2}_0
\sech^2(\kappa^*_0 x_{R1}^*) - {\rm i}g] = - 2 {\rm i}g$, where use has been made of
(\ref{eq:C1}) and (\ref{eq:C2}). Similarly, from equations (\ref{eq:energy}) and
(\ref{eq:relation-1}) it follows that $\Delta V^{(-)}(\pm l) = {\rm i} g$.\par
%
%
This confirms that $V^{(-)}(x)$ has the same discontinuities as $V^{(+)}(x)$. However,
when we compare the jumps of the former with those of the latter resulting from
definition (\ref{eq:V-i}), we find
\be
  \Delta V^{(-)}(x^{(i)}) = - \Delta V^{(+)}(x^{(i)}), \qquad i=1,2,3.  \label{eq:jump}
\ee
Such a behaviour can be traced back to the superpotential, which turns out to be a
continuous function of $x$ on $(-L, +L)$, in contrast with its derivative, which is
discontinuous at $x^{(i)}$, $i=1$, 2, 3. The third relation in (\ref{eq:definition}) then
immediately leads to (\ref{eq:jump}).\par  
%
\section{Conclusion}

Under the simplest assumption of unbroken SUSY, we have shown that for the weakly
non-Hermitian square well with three discontinuities at $x = -l$, 0 and $l$, the SUSY 
partners $H^{(\pm)}$ are both non-Hermitian and $\cal PT$-symmetric.
Moreover, the partner potential $V^{(-)}(x)$ has the same three
discontinuities  as $V^{(+)}(x)$.\par
%
%
It should be noted that in the two limiting cases $l \to 0$ and $l \to L$, our results give
back those relative to the real square well \cite{sukumar} and to the $\cal
PT$-symmetric square well with a single discontinuity \cite{CQ}, respectively.\par
%
%
It is conjectured that as for the strongly non-Hermitian square well with a single
discontinuity at $x=0$~\cite{sqw}, a charge-conjugation operator $\cal C$ \cite{BBJ} may
be constructed in a specific form differing from the unit operator mostly in a
finite-dimensional subspace of the Hilbert space \cite{Batal}.
This is one of the most important merits of all the square-well
models with $L < \infty$. It seems to open a new inspiration for a
direct physical applicability of non-Hermitian models whenever
their spectrum remains real. \par
%
%
\bigskip

\noindent
{\small {\bf Acknowledgements}. Work partially supported by AS CR
(GA grant Nr. A1048302 and IRP AV0Z10480505). C. Quesne is a
Research Director, National Fund for Scientific Research (FNRS),
Belgium.}\par
%
%

\end{document}